\documentclass[a4paper]{article}
\usepackage{amsmath,amssymb,comment,graphicx}

\newcommand{\A}{\mathcal{A}}
\newcommand{\B}{\mathcal{B}}
\newcommand{\C}{\mathcal{C}}

\newcommand{\E}{\mathcal{E}}
\newcommand{\F}{\mathcal{F}}
\newcommand{\G}{\mathcal{G}}
\renewcommand{\H}{\mathcal{H}}

\newcommand{\OO}{\mathcal{O}}

\newcommand{\para}[1]{\smallskip\noindent{\bf #1}}

\newcommand{\ext}[1]{{\bf Ext}({#1})}
\newcommand{\cyc}[1]{{\bf Cyc_a}({#1})}
\newcommand{\cyca}[1]{{\bf Cyc_a}({#1})}
\newcommand{\cycf}[1]{{\bf Cyc_f}({#1})}
\newcommand{\cycg}[1]{{\bf Cyc_g}({#1})}

\newcommand{\PP}[1]{\mathbb{P}_{#1}}

\newtheorem{thm}{Theorem}
\newtheorem{lem}[thm]{Lemma}

\newtheorem{cor}[thm]{Corollary}

\newenvironment{pf}[1][Proof:]{\begin{trivlist}
\item[\hskip \labelsep {\em #1}]}{\hfill$\Box$\end{trivlist}}

\newtheorem{pro-example}{Example}
\newenvironment{example}{\begin{pro-example}\rm}{\hfill$\diamond$\end{pro-example}}

\newcommand{\qed}{ }

\begin{document}

\title{Fast Synchronization of Random Automata\footnote{This work is supported by the French National Agency (ANR) through ANR-10-LABX-58, through ANR-JCJC-12-JS02-012-01 and through ANR-2010-BLAN-0204.}}

\author{Cyril Nicaud \\LIGM, Universit\'e Paris-Est \& CNRS,\\ 5 bd Descartes, Champs-sur-Marne, 77454 Marne-la-Vall\'ee Cedex 2, France\\\texttt{nicaud@univ-mlv.fr}}

\maketitle 

\begin{abstract}
A synchronizing word for an automaton is a word that brings  that automaton
into one and the same state, regardless of the starting position.
\v{C}ern\'y conjectured in 1964 that if a $n$-state deterministic automaton
has a synchronizing word, then it has a synchronizing word of 
size at most $(n-1)^2$. Berlinkov recently made a breakthrough in the probabilistic
analysis of  synchronization by proving that with high probability, an automaton has a synchronizing word.
In this article, we prove that with high probability an automaton admits a synchronizing word
of length smaller than $n^{1+\epsilon}$, and therefore that
the \v{C}ern\'y conjecture holds with high probability.
 \end{abstract}


\section{Introduction}
A \emph{synchronizing word} (or a \emph{reset word}) for an automaton is a word 
that brings that automaton into one and the same state, regardless of the starting position.
This notion,  first formalized by 
\v{C}ern\'y in the sixties, arises naturally in automata theory and its
extensions, and plays an important role in several application areas~\cite{volkov08}.
Perhaps one of the reasons synchronizing automata are still intensively studied in 
theoretical computer science is a beautiful question asked by \v{C}ern\'y~\cite{cerny} back in 1964:
``\emph{Does every synchronizing $n$-state automaton admits a synchronizing word of length at most $(n-1)^2$?}''
The bound of $(n-1)^2$, as shown by \v{C}ern\'y, is best possible.
This question, known as \emph{\v{C}ern\'y's conjecture}, is one of the most famous conjectures 
in automata theory. 
Though established for important subclasses of automata,  
\v{C}ern\'y's conjecture remains open in the general case. The best  known upper bound,
established in the early eighties~\cite{pin,frankl},
is $\frac16(n^3-n)$.
For a more detailed account on \v{C}ern\'y's conjecture,
we refer the interested reader to Volkov's article~\cite{volkov08}.

\para{Probabilistic \v{C}ern\'y conjecture.} 
Considering \v{C}ern\'y's conjecture from  a probabilistic point
of view is natural (see for instance~\cite{cameron}), and leads to the following questions: 

\null\hfill
\begin{minipage}{.95\textwidth}
{\bf Question 1:} Is a random  automaton synchronizing \emph{with high probability}?\\
{\bf Question 2:} Does a synchronizing $n$-state automaton  admits a synchronizing word of length at most  $(n-1)^2$ \emph{with high probability}?
\end{minipage}
\hfill\null

\medskip

\noindent Here, \emph{with high probability}  
means ``with probability that tends to $1$ as $n$ goes to infinity''. 

Berlinkov recently made a breakthrough by giving a positive answer to Question~1~\cite{berlinkov}: he proved that the probability 
that a random automaton is not synchronizing is in $\OO(n^{-\frac12 |A|})$, for an alphabet
$A$ with at least two letters.

Question~2 can be simulated and experimental
evidence suggests that most automata are synchronized by a short
synchronizing word, of length sublinear in the number of states. Note that
simulating the second question is nontrivial, as finding the shortest reset 
word is hard~\cite{Olschewski}; the best experimental results we are aware of
were obtained by Kisielewicz, Kowalski, and Szykula~\cite{Kisielewicz}.

\para{Our results.}
In this paper we give a positive answer to Question~2 
when the automaton is chosen uniformly among
deterministic and complete $n$-state automata on an alphabet with at least two letters.
More precisely, we show that for any $\epsilon>0$, the
probability that a random $n$-state automaton has a synchronizing 
word of length smaller than $n^{1+\epsilon}$ tends to $1$ when $n$ goes to infinity. 
Of course, an immediate consequence is that if \v{C}ern\'y's conjecture is false, a counterexample will hardly be found by mere uniform random exploration.

Our proof also gives another way to show that automata are synchronizing with high probability, 
based on a completely different
method: Berlinkov used advanced properties on the highest tree in a random mapping to
study the \emph{stable pairs} of the automaton, where we directly build words that iteratively shrink the set of states,
using only basic discrete probabilities and variations on the probabilistic pigeonhole principle\footnote{Also known as the Birthday Paradox.}. The proof proposed by Berlinkov is arguably more complicated but also more 
precise since it gives the error term in $\OO(n^{-\frac12|A|})$ for the probability of not being synchronizing\footnote{Knowing the probability of not being synchronizing  is important in many situations, especially for the average case analysis of algorithms as illustrated in
the conclusions of~\cite{berlinkov}.}. There is little hope that the method presented below can be used to achieve such a precise estimation of the number of non-synchronizing automata.

\section{Definitions and notations}\label{sec:def}
\para{Basic notations.} For any integer $n\geq 1$, let
$[n]=\{1,\ldots,n\}$ be the set of integer from $1$ to $n$. The cardinality of
a finite set $E$ is denoted by $|E|$.

\para{Automata.} 
Let $A$ be a finite alphabet, a \emph{deterministic automaton} on $A$ 
is a pair $(Q,\delta)$, where $Q$ is a finite set of \emph{states} and $\delta$ is the \emph{transition function},
a (possibly partial) mapping from $Q\times A$ to $Q$. If $p,q\in Q$ and $a\in A$ are such that $\delta(p,a)=q$,
then $(p,a,q)$ is the \emph{transition} from $p$ to $q$ labelled by $a$, and is denoted by 
$p\xrightarrow{a}q$. It is an \emph{$a$-transition} outgoing from $p$. 

In this article, we are not interested in initial and final states since they do not
matter for the synchronization. We will also focus on deterministic automata only, 
and therefore, throughout the article,
we will simply call ``automaton'' a deterministic automaton with no initial and final states.

An automaton $\A=(Q,\delta)$ on $A$ is classically seen as a labelled 
directed graph of set of vertices $Q$ and whose edges are the transitions of $\A$.

An automaton is \emph{complete} when its transition function is a total function and \emph{incomplete}
otherwise. The transition function is extended inductively to $Q\times A^*$  by setting 
$\delta(p,\varepsilon)=p$ for every $p\in Q$ and, for every $u\in A^{*}$, 
$\delta(p,ua)=\delta(\delta(p,a),u)$ when everything is defined, and undefined otherwise. If $u\in A^*$, we denote
by $\delta_u$ the (possibly partial) function from $Q$ to $Q$ defined by $\delta_u(p) = \delta(p,u)$, for all $p\in Q$.

If $\A=(Q,\delta)$ is an automaton on $A$, an \emph{extension} of $\A$ is an automaton $\B=(Q,\lambda)$ on $A$ such that
for all $p\in Q$ and all $a\in A$, if $\delta(p,a)$ is defined then $\lambda(p,a)=\delta(p,a)$. The automaton
$\B$ is therefore obtained from $\A$ by adding some transitions. We denote by $\ext{\A}$ the set of
all the extensions of an automaton $\A$. If $\H$ is a set of automata, we denote by $\ext{\H}$ the union of all the $\ext{\A}$ for $\A\in\H$.

\para{Synchronization.}
Let $\A$ be an automaton on $A$. Two states $p$ and $q$ of $\A$ are \emph{synchronized} by the
word $w\in A^*$ when both $\delta_w(p)$ and $\delta_w(q)$ exist and are equal. 

A \emph{synchronizing word} for an automaton $\A=(Q,\delta)$ is a word $w\in A^*$ such that
$\delta_w$ is a constant map: there exists a state $r\in Q$ such that for every $p$ in $Q$,
$\delta_w(p)=r$. An automaton that has a synchronizing word is said to be \emph{synchronizing}.

\para{Mappings.}
A \emph{mapping} on a set $E$ is a total function from $E$ to $E$. When $E$ is finite,
a mapping $f$ on $E$ can be seen as a directed graph with an edge $i\rightarrow j$ whenever
$f(i)=j$.  An example of such a graph is depicted in Figure~\ref{fig:a} page~\pageref{fig:a}.

If $f$ is a mapping on $E$, $x\in E$ is a \emph{cyclic point} of $f$ (or \emph{$f$-cyclic point} when there are several mappings) when there exists an integer $i>0$ such that $f^i(x)=x$. 
In the sequel, $E$ will often be the set of states of an automaton, and we will therefore use the term
``state'' instead of ``point'': $f$-cyclic state, \ldots

If $f$ is a mapping on $E$ and $x\in E$, the \emph{height} of $x$ is the smallest $i\geq 0$
such that $f^i(x)$ is a cyclic point. The height of a cyclic point is therefore $0$. The \emph{height} of
a mapping on $E$ is the maximal height of the elements of $E$. The mapping depicted in Figure~\ref{fig:a} page~\pageref{fig:a} has height $3$,
and the maximal height is reached by $9$.

\para{Probabilities.}
Let $(E,s)$ be a pair where  $E$ is a set and $s$ is a \emph{size function} $s$ from $E$ to
$\mathbb{Z}_{\geq 0}$. The pair $(E,s)$ is a combinatorial set\footnote{The size is often clear in the context (number of nodes in a tree, ...)
and can be omitted.} when for every integer $n\geq 0$,
the set $E_{n}$ of size-$n$ elements of $E$ is finite. 
 To simplify the definitions,
we also assume that $E_n\neq\emptyset$ for every $n\geq 1$, which will always be the case in the following. 
Let $(\mathbb{P}_n)_{n\geq 1}$ be a sequence of total functions such that for each $n\geq 1$, $\mathbb{P}_n$ is a probability 
on $E_n$.  We say that a property $P$ holds \emph{with high probability} (\emph{whp}) for $(\mathbb{P}_n)_{n\geq 1}$ when
$\mathbb{P}_n[P\text{ holds}]\rightarrow 1$ as $n\rightarrow \infty$.

We will often consider the \emph{uniform distribution} on $E$, which is the sequence $(\mathbb{P}_n)_{n\geq 1}$
defined by $\mathbb{P}_n[\{e\}] = \frac1{|E_n|}$ for any $e$ in $E_n$: A sentence like ``property $P$ holds \emph{whp} for the uniform distribution 
on $E$'' therefore means that the probability that $P$ holds tends to $1$ as $n$ tends to infinity, 
when for each $n$ we consider the uniform distribution on $E_n$.
The reader is referred to~\cite{FSBook} for more information on combinatorial probabilistic
models.

\para{Random mappings and random $p$-mappings.}
A \emph{random mapping} of size $n\geq 1$ is a mapping on $[n]$ taken with the uniform distribution. If
$p$ is a probability mass function on $[n]$, a random $p$-mapping is the distribution on the mappings on $[n]$ such that
the probability of a mapping $f$ is $\prod_{i\in[n]}p(f(i))$: the image of each $i\in[n]$ is chosen independently
following the probability $p$. 

A result stated as ``a random $p$-mapping satisfies property $P$ \emph{whp}'' means that for \emph{any} sequence
$(p_n)_{n\geq 1}$, where $p_n$ is a probability on $[n]$, the probability that a $p_n$-random mapping on $[n]$
satisfies $P$ tends to $1$ as $n$ tends to infinity. It is therefore a strong result that does not depend
on the choice of $(p_n)_{n\geq 1}$.

\para{Random automata.}
In the sequel, the set of states of an $n$-state automaton will always be $[n]$. With this condition, there are
exactly $n^{|A|n}$ complete automata with $n$ states on $|A|$, and we are therefore interested in the uniform distribution where each size-$n$
complete automaton has probability $n^{-|A|n}$. Note that one can also see this distribution as drawing uniformly at random
and independently in $[n]$ the image of each $\delta(p,a)$, for all $p\in[n]$ and $a\in A$. These alternative way to look
at random automata will be widely used in the sequel, especially in the following way: for a fixed incomplete automaton
$\A$ with $n$ states, the uniform distribution on complete automata of $\ext{\A}$ can be seen as setting uniformly 
at random and independently in $[n]$ the transitions that are undefined in $\A$.

\section{Preliminary classical results}\label{sec:prelim}
In this  section, we recall some classical results that will be useful in sequel. 
Though elementary, these results are the main ingredients of this article. The
proofs are not new but given for completeness.
  
We start with the following property for synchronizing automata: an automaton is synchronizing
if and only if every pair of states can be synchronized.
\begin{lem}\label{lm:pairs}
Let $\A$ be an $n$-state automaton and $\ell$ be a non-negative integer. If for every pair of
states $(p,q)$ in $\A$ there exists a word $u$ of length at most $\ell$ such that
$\delta_u(p)=\delta_u(q)$, then $\A$ admits a synchronizing word of length at most
$\ell(n-1)$.
\end{lem} 

\begin{pf}
Assume we successfully synchronized $i$ pairwise distinct states $q_1$, \ldots $q_i$
using a word $u$ of length smaller than or equal to $\ell(i-1)$: for all $j,k\in\{1,\ldots i\}$, 
$\delta_u(x_j)= \delta_u(x_k)$. Let $x_{i+1}$ be a state distinct from $x_1$,\ldots,$x_i$ and 
let $v$ be a word of length at most $\ell$
that synchronizes $\delta_u(x_1)$ and $\delta_u(x_{i+1})$. Then the word $uv$ synchronizes $x_1$,\ldots ,$x_{i+1}$
and has length at most $\ell\cdot i$. The result follows by induction.\qed
\end{pf}

 Random mappings and random $p$-mappings have been studied intensively
in the literature~\cite{harris,rm,kolcin}, using probabilistic techniques or methods from analytic combinatorics.
In this section, we only recall basic properties of the typical number of cyclic points and of the
typical height of a random $p$-mapping. This can be achieved using variations on the probabilistic pigeonhole principle only;
more advanced techniques can be used to obtain more precise 
statements\footnote{For instance, limit distributions of some parameters~\cite{FSBook} or even a notion of continuous limit for random mappings~\cite{aldous}.}, 
but we will only need the following results in the sequel\footnote{The bound are not tight, we choose them for  readability.}.

\begin{lem}\label{lm:p cyclic points}
Let $\epsilon>0$ be a fixed real number.
For $n$ large enough, the probability that a random $p$-mapping of size $n$ 
has more than $n^{\frac12+\epsilon}$ cyclic points or that its height is greater than $n^{\frac12+\epsilon}$ is
at most $\exp(- n^{\epsilon})$.
\end{lem}

The proof of Lemma~\ref{lm:p cyclic points} consists in two steps. It is first established for uniform random mappings
then extended to general $p$-random mappings using the following technical folklore lemma.

\begin{lem}\label{lm: bounding sym sums}
Let $n$ and $\ell$ be two positive integers such that $\ell\leq n$.
Let $(E,\leq)$ be a totally ordered finite set of cardinality $n$.
Let $f$ be a map from $E$ to $\mathbb{R}_{\geq 0}$, and denote by $s$ the sum
of the images by $f$: $s=\sum_{x\in E}f(x)$. The following result holds:
\[
\sum_{x_1<x_2<\ldots <x_\ell} f(x_1)f(x_2)\cdots f(x_\ell) \leq \binom{n}{\ell} \left(\frac{s}{n}\right)^{\ell},
\]
where the sum range over all increasing $\ell$-tuples of elements of $E$.
The sum on the left is therefore maximal when  $f(x) = \frac{s}{n}$, for every $x\in E$.
\end{lem} 

\begin{pf}
Let $\nu(f)$ denote the number of elements $x\in E$ such that $f(x)$ is different from $\frac{s}{n}$:
\[
\nu(f) = \left| \left\{x\in E:\ f(x)\neq\cfrac{s}{n} \right\}\right|.
\]
We prove  by induction on the value of $\nu$ that every map from $E$ to $\mathbb{R}_{\geq0}$ that sums up
to $s$ satisfies the inequality stated in the lemma.

\noindent $\blacktriangleright$ If $\nu(f)=0$ then $f(x)=\frac{s}{n}$ for every $x\in E$. Thus, for any $\ell$-tuple $(x_1,\ldots,x_\ell)$
we have that 
\[
f(x_1)\cdots f(x_\ell) = \left(\frac{s}{n}\right)^\ell,
\]
and therefore, since there are $\binom{n}{\ell}$ such increasing sequences,
\[
\sum_{x_1<\ldots<x_\ell} f(x_1)\cdots f(x_\ell) = \binom{n}{\ell}\left(\frac{c}{n}\right)^\ell.
\]

\noindent $\blacktriangleright$  Assume now that $\nu(f)\neq 0$. 
For the induction step, we build another map $g$, starting from $f$, such that $g$ sums up to $s$, $\nu(g)<\nu(f)$ and 
\[
\sum_{x_1<\ldots<x_\ell} f(x_1)\cdots f(x_\ell)\leq
\sum_{x_1<\ldots<x_\ell} g(x_1)\cdots g(x_\ell).
\]
Let $y\in E$ be an element such that $|f(y)-\frac{s}n|$ is minimal
amongst the $y$'s such that $f(y)\neq\frac{s}n$. We assume that $f(y)-\frac{s}n > 0$, the proof if almost the same if $f(y)-\frac{s}n < 0$.
Since $f(y)>\frac{s}n$ and $\sum_{x\in E}f(x) = s$, there exists an element $z\neq y$ such that $f(z)<\frac{s}n$.
Consider the new map $g$ obtained from $f$ by changing the value of $y$ and $z$ the following way:
\[
\begin{cases}
g(x) = f(x) \text{ if }x\neq y\text{ and }x\neq z, \\
g(y) = \frac{s}n, \\
g(z) = f(z) + f(y) - \frac{s}n.
\end{cases}
\]
It is direct to verify that $g$ is always non-negative and sums up to $s$. Moreover, by construction
$\nu(g)<\nu(f)$. We claim that
\begin{equation}\label{ax:inequality}
\sum_{x_1<\ldots<x_\ell} f(x_1)\cdots f(x_\ell) \leq \sum_{x_1<\ldots<x_\ell} g(x_1)\cdots g(x_\ell).
\end{equation}
To prove this inequality, we consider three cases:

$\bullet$ If an increasing $\ell$-tuple $(x_1,\ldots,x_\ell)$ does not contain
$y$ nor $z$, then we have $g(x_1)\cdots g(x_\ell) = f(x_1)\cdots f(x_\ell)$. 

$\bullet$ We sum the contributions of tuples containing exactly one of $y$ or $z$: if
$\{x_1,\ldots,x_{\ell-1}\}$ are $\ell-1$ elements of $[n]\setminus\{y,z\}$ we have
\begin{align*}
g(x_1)\cdots g(x_{\ell-1})\,g(y) &+ g(x_1)\ldots g(x_{\ell-1})\,g(z) \\
&  = g(x_1)\cdots g(x_{\ell-1})\,\left(g(y) + g(z) \right) \\
& = f(x_1)\cdots f(x_{\ell-1})\,\left(\frac{s}n  + f(z) + f(y) - \frac{s}n\right)\\
& = f(x_1)\cdots f(x_{\ell-1})\,f(y) + f(x_1)\cdots f(x_{\ell-1})\,f(z). 
\end{align*}
Hence the contributions of such tuples globally do not change the value of the sum when switching from
$f$ to $g$.

$\bullet$ If both $y$ and $z$ are in the tuple, then
\[
\begin{cases}
f(x_1)\cdots f(x_\ell) = f(y)f(z)\prod_{\substack{x_i\neq y\\x_i\neq z}} f(x_i) \\
g(x_1)\cdots g(x_\ell) = g(y)g(z)\prod_{\substack{x_i\neq y\\x_i\neq z}} g(x_i) 
 = g(y)g(z)\prod_{\substack{x_i\neq y\\x_i\neq z}} f(x_i).
\end{cases}
\]
Let $\alpha$ and $\beta$ be the two positive real numbers defined by $\alpha = f(y)-\frac{s}n$ and
$\beta = \frac{s}n-f(z)$. We have
\[
g(z)g(y)=\frac{s}n\left(f(z) + f(y) - \frac{s}n\right) = \frac{s}n\left(\frac{s}n + \alpha-\beta\right) =
\frac{s^2}{n^2} + \frac{s(\alpha-\beta)}n,
\]
whereas
\[
f(y)\,f(z) = \left(\frac{s}n+\alpha\right)\left(\frac{s}n-\beta\right) = \frac{s^2}{n^2} + \frac{s(\alpha-\beta)}n
-\alpha\beta,
\]
therefore $f(y)\,f(z)\leq g(y)\,g(z)$ and $f(x_1)\cdots f(x_\ell)\leq g(x_1)\cdots g(x_\ell)$ for such a tuple.

This proves Equation~\eqref{ax:inequality} and concludes the proof by induction on the value of $\nu$.\qed
\end{pf}

\begin{pf}[Proof of Lemma~\ref{lm:p cyclic points}:]
We start with the uniform case.

\noindent$\blacktriangleright$ Number of cyclic points (uniform case):
For any integer $\ell$ such that $1\leq \ell \leq n$, the probability
that there is a cyclic part of size $\ell$ in a uniform random mapping is at most
$P(n,\ell)=\binom{n}{\ell} \ell !\ n^{-\ell}$,
since we need to choose the $\ell$ elements that form the cyclic part, the way they are mapped to
each other, and this forces $\ell$ images of the map which are correctly set with probability
$\frac1n$ each. This is an upper bound since we do not prevent the formation of other cycles in our counting. Moreover,
\begin{align*}
P(n,\ell) & = \frac{n!}{(n-\ell)!} n^{-\ell} = \left(1-\frac1n\right)\left(1-\frac2n\right)\cdots\left(1-\frac{\ell-1}{n }\right) = \prod_{i=1}^{\ell-1}\left(1-\frac{i}n\right) \\
& \leq \exp\left(-\sum_{i=1}^{\ell-1}\frac{i}n\right)  \leq \exp\left(-\frac{\ell(\ell-1)}{2n}\right).
\end{align*}
Hence, the probability that there is a cyclic part of length greater than or equal to 
$n^{\frac12+\epsilon}$ is at most, for $n$ large enough,
\[
\sum_{\ell = \lceil n^{\frac12+\epsilon}\rceil}^{n}P(n,\ell) \leq n\cdot \exp\left(-\frac{\lceil n^{1/2+\epsilon}\rceil(\lceil n^{1/2+\epsilon}\rceil-1)}{2n}\right)
\leq \frac12\exp\left(- n^{\epsilon}\right).
\] 

\noindent$\blacktriangleright$ Height (uniform case): Consider an element
$i\in[n]$. For any integer $\ell$ such that $0< \ell < n$, the probability that a uniform random mapping
$f$ on $[n]$ is such that $f(i)$, $f^2(i) = f(f(i))$, \ldots,
$f^{\ell}(i)$ are all distinct is classically
\[
\left(1-\frac1n\right)\left(1-\frac2n\right) \cdots \left(1-\frac{\ell-1}n\right) = P(n,\ell).
\]
We can therefore use the previous computations. If $f$ has height greater than or equal to $\ell$, then 
there exists a $i$ with more than $\ell$ distinct iterates. Hence, using the union bound by 
summing the contribution of all $i\in[n]$, we get that the probability that $f$ has height greater than
$\lceil n^{1/2+\epsilon} \rceil$ is at most, for $n$ large enough,
\[
n\cdot P(n,\lceil n^{1/2+\epsilon}\rceil) \leq n\cdot\exp\left(-\frac{\lceil n^{1/2+\epsilon}\rceil(\lceil n^{1/2+\epsilon}\rceil-1)}{2n}\right)
\leq \frac12\exp\left(-n^{\epsilon}\right).
\]
This concludes the proof for the uniform case.

\medskip

\noindent We now consider the case of  random $p$-mappings.

\noindent$\blacktriangleright$ Number of cyclic points (non-uniform case):
We start as in the proof for the uniform case. The difference is that if the points involved
in the cyclic part of length $\ell$ are $x_1$, $x_2$, \ldots, $x_\ell$ then the upper 
bound for the probability is not $\ell!\, n^{-\ell}$ anymore\footnote{The $\ell!$ term is for counting the number
of way to map bijectively the $x_i$'s to themselves, forming the cyclic part (or a part of it).} but
\[
P_n(x_1,\ldots,x_\ell) = \ell!\ p(x_1)p(x_2)\cdots p(x_\ell).
\]
If we sum this quantity on all possible $\ell$-subsets of $[n]$ we obtain an upper bound of
\[
P_n(\ell) = \ell!\ \sum_{1\leq x_1<x_2<\ldots<x_\ell\leq n} p(x_1)p(x_2)\cdots p(x_\ell).
\]
At this point we can apply Lemma~\ref{lm: bounding sym sums} with $f=p$, $E=[n]$ and $s=1$ 
to obtain a uniform bound for $P_n(\ell)$:
\[
P_n(\ell) \leq \ell! \binom{n}{\ell} n^{-\ell},
\]
and this is the same bound as for the uniform case, yielding the same result.

\noindent$\blacktriangleright$ Height (non-uniform case):
Similarly, we start with the same idea as in the proof for the uniform case. 
Fix some $x\in[n]$. Let $(x_1,\ldots,x_\ell)$ be a  $\ell$-tuple of distinct elements in $[n]\setminus\{x\}$.
The probability that a map $f$ is such that  $x_{i} = f^i(x)$, for all $1\leq i \leq \ell$, is simply
$p(x_1)p(x_2)\cdots p(x_{\ell})$. Hence  the probability that $x$
has $\ell$ distinct iterates that are different from $x$ when applying $f$ is
$p(x_1)p(x_2)\cdots p(x_{\ell})$ where $(x_1,\ldots,x_\ell)$ ranges over all $\ell$-tuples
of pairwise distinct elements of $[n]\setminus\{x\}$. We obtain an upper bound by allowing one of
the $x_i$'s to be equal to $x$, which simplifies the writing; the bound is:
\[
\ell!\ \sum_{1\leq x_1<x_2<\ldots<x_\ell\leq n} p(x_1)p(x_2)\cdots p(x_\ell), 
\]
since there are $\ell!$ ways to permute the $x_{i}$'s. 
This is the same quantity as $P_{n}(\ell)$ for
the number of cyclic points just above, yielding the same result and concluding the proof.\qed
\end{pf}

\section{Main Result}\label{sec:proof}
The main result of this article is the following theorem.

\begin{thm}\label{thm:main}
Let $\epsilon$ be a positive real number smaller than $\frac18$, and let $A$ be an alphabet
with at least two letters.  For the uniform distribution, an $n$-state deterministic and complete automaton 
on $A$  admits a synchronizing word of length smaller than $n^{1+\epsilon}$ with high probability. More precisely,
the probability it has no such word is in $\OO(n^{-\frac18+\epsilon})$.
\end{thm} 

The statement does not hold for alphabets with only one letter, since there 
are cycles of length greater than $1$ in a random mapping \emph{whp}~\cite{rm}: two distinct states
in such a cycle cannot be synchronized.

As a  consequence of Theorem~\ref{thm:main},  a random deterministic and complete automaton is
synchronizing \emph{whp}; our proof therefore constitutes an alternative proof of~\cite{berlinkov} for 
that property. Our statement is weaker since Berlinkov also obtained bounds  in $\OO(n^{-\frac12|A|})$ for the error 
term (the number of automata that are not synchronizing), which is tight for two-letter alphabets. On the other hand, it is arguably 
more elementary as we mostly rely on Lemma~\ref{lm:p cyclic points} and some basic discrete probabilities; in any cases, we hope to shed a new light on the reasons why automata are often synchronizing.

If we consider the uniform distribution on synchronizing automata,
we directly obtain that there  exists a small synchronizing word \emph{whp}, yielding the following corollary.
\begin{cor}\label{cor:main}
For the uniform distribution on synchronizing deterministic and complete automata on an alphabet
with at least two letters, \v{C}ern\'y's conjecture holds with high probability.
\end{cor}

\bigskip

\noindent We prove Theorem~\ref{thm:main} in two main steps:

\noindent$\blacktriangleright$ We first construct a word $w_n\in\{a,b\}^*$ such that the image of $\delta_{w_n}$
for a random $n$-state automaton  has  size at most $n^{1/8+4\epsilon}$ \emph{whp}. This is done by building a set $\G_n$ of
incomplete automata, \emph{all} of which have the desired property, and showing that a random $n$-state 
automaton  extends an element of $\G_n$ \emph{whp}. Roughly speaking,  $\G_n$ and $w_n$ are built 
by three consecutive applications of Lemma~\ref{lm:p cyclic points}, starting with incomplete automata
with only $a$-transitions, which we then augment by $b$-transitions in two rounds.

\medskip

\noindent$\blacktriangleright$ It remains to synchronize those $n^{1/8+4\epsilon}$ states. 
This is done by showing that for a
random automaton that extends an element of $\G_n$,  any two among those $n^{1/8+4\epsilon}$
states can  be synchronized by a word of the form $b^iw_n$ \emph{whp}, with $i\leq n^{1/8+5\epsilon}$.
Lemma~\ref{lm:pairs} is then used to combine these words, and also $w_n$, into a synchronizing word for that automaton.


\bigskip
The remainder of this section is devoted to a more detailed proof of Theorem~\ref{thm:main}. For the
presentation, we will follow an idea used by Karp in his article on random direct graphs~\cite{Karp90}: we
start from an automaton with no transition, then add new random 
transitions during at each step
 of the construction, progressively improving the synchronization.

\bigskip

From now on, we fix a real $\epsilon>0$ small enough\footnote{To simplify the writing,
we will prove that  there is a synchronizing word
of length $\OO(n^{1+11\epsilon})$ \emph{whp}, and then get the statement of Theorem~\ref{thm:main} by changing $\epsilon$
into $\epsilon/13$.}. Since it is clearly sufficient to establish
the result for a two-letter alphabet, we consider that $A=\{a,b\}$ in the sequel.

\subsection{Generating the $a$-transitions}
The first step consists in generating all the $a$-transitions. This forms a mapping
for $\delta_a$ that follows the uniformly distribution on size-$n$ mappings. 
We can therefore apply  Lemma~\ref{lm:p cyclic points}, and obtain that words of the form $a^i$
can already be used to reduce the number of states to be synchronized.

Let $\alpha_n=\lfloor n^{\frac12 + \epsilon}\rfloor$ and 
let $\E_n$ denote the set of incomplete automata $\A$ with $n$ states such that:
\begin{enumerate}
\item the defined transitions of $\A$ are exactly its $a$-transitions;
\item the action $\delta_a$ of $a$  has at most $\alpha_n$ cyclic states;
\item the height of $\delta_a$ is at most $\alpha_n$. 
\end{enumerate}
An example of an element of $\E_n$ is given below.

\begin{example}\label{ex:a}
 Let $\A$ be a mapping with 18 states, which has only $a$-transitions and
such that $\delta_{a}$ is the mapping of Figure~\ref{fig:a}.
\begin{figure}[h]
\centering
\includegraphics{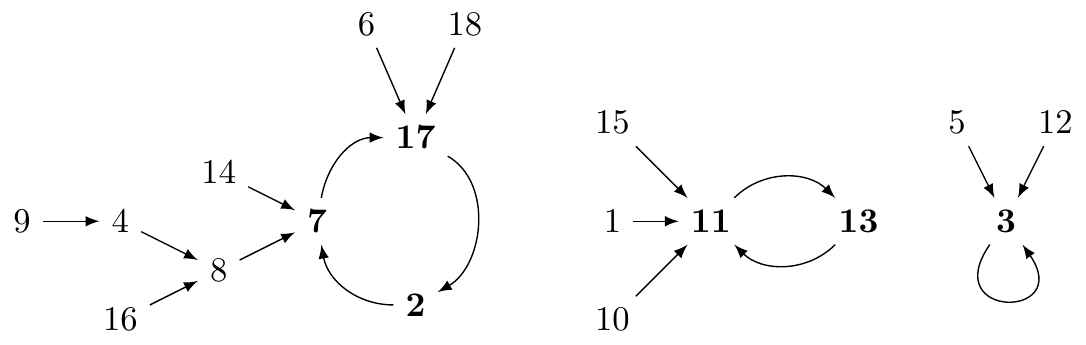}
\caption{This mapping represents the action of $a$ in the automaton $\A$.\label{fig:a}}
\end{figure}

The set $\cyc{\A}$ is made of the bold labels
$\{{\bf 2},{\bf 3},{\bf 7},{\bf 11},{\bf 13},{\bf 17}\}$. Assume that for our $\epsilon$ we have $\alpha_{18}=6$, then $u_n=aaaaaa$ is used to start the synchronization:
\[
\begin{array}{ccc}
\{2,8,14\} \xrightarrow{u_n} {\textbf{2}};
 &
\{3, 5, 12\} \xrightarrow{u_n} {\textbf{3}};
&
\{6, 7, 9, 18\} \xrightarrow{u_n} {\textbf{7}};
\\
\{11\} \xrightarrow{u_n} {\textbf{11}};
&
\{1, 10, 13, 15\} \xrightarrow{u_n} {\textbf{13}};
&
\{4, 16, 17\} \xrightarrow{u_n} {\textbf{17}}.
\end{array}
\]
Since there are  $6\leq\alpha_{18}$ cyclic states and since the height of the mapping is $3\leq\alpha_{18}$, $\A$
is in $\E_n$.
\end{example}

By independency, the action of letter $a$ in a uniform random complete automaton is exactly
a uniform random mapping, yielding the following consequence of Lemma~\ref{lm:p cyclic points}.

\begin{lem}\label{lm:Cn}
A random complete automaton with $n$ states extends an element of $\E_n$ \emph{whp}. More precisely, the probability
that such an automaton does not extend an element of $\E_n$ is at most
$\exp(-n^{\epsilon})$.
\end{lem}

For any automaton $\A$ whose $a$-transitions are all defined, 
let $\cyc{\A}$ denote its set of $\delta_a$-cyclic states; they also 
are the $\delta_a$-cyclic states of any automaton that extends $\A$. 

Let $u_n=a^{\alpha_n}$. By Lemma~\ref{lm:Cn}, we can 
already start the synchronization using $u_{n}$, since \emph{whp} 
the image of the set of states $[n]$ by $\delta_{u_n}$ is  
included in $\cyca{\A}$, which has size at most $\alpha_n$. In the sequel, we therefore
work on synchronizing the elements of $\cyca{\A}$.

\subsection{Adding random $b$-transitions that start from the $\delta_a$-cyclic states}\label{sec:b}
Let $\A$ be a fixed element of $\E_n$. We are now working on $\ext{\A}$ and we consider the
process of adding a random $b$-transition starting from every state of $\cyca{\A}$. 

Let $\B\in\ext{\A}$ be an automaton obtained this way 
and let $f_\B$ denote the restriction of $\delta_{bu_n}$ to $\cyca{\A}$. It 
is a total map, since all the needed $b$-transitions are defined. Moreover, the image of $f_\B$ 
is included in $\cyca{\A}$, as $f_\B(x)=\delta_{bu_n}(x)=\delta_{u_n}(\delta_b(x))$, for every $x\in\cyca{\A}$. Hence $f_\B$
is a total map from $\cyca{\A}$ to itself. 

\begin{example}\label{ex:b}
This is the automaton of Example~\ref{ex:a}, where the  $b$-transitions that
start from the elements of $\cyca{\A}$ have been added (in bold):

\centerline{\includegraphics{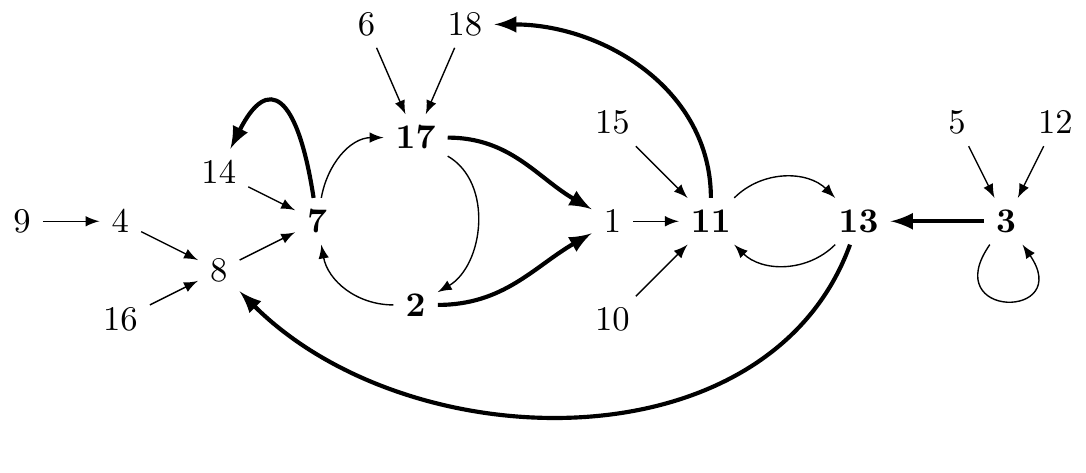}}

\noindent Below is depicted the map $f_\B$, which is the restriction of $\delta_{bu_n}$ to $\cyca{\A}$;
an edge ${\bf p}=x\Longrightarrow {\bf q}$ means that $\delta_b(p)=x$ and
$\delta_{u_n}(x)=q$, so that $f_\B(p)=q$:

\centerline{\includegraphics{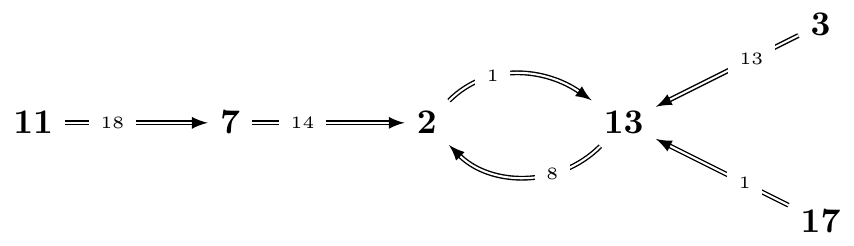}}
\end{example}

From a probabilistic point of view, if we fix $\A$ and build $\B$ 
by adding uniformly at random and independently the  $b$-transitions that start from the states of $\cyca{\A}$, 
the induced distribution for the mapping $f_\B$ is usually not the uniform distribution on mappings of $\cyca{\A}$. 
More precisely, for any $q\in\cyca{\A}$ the probability that 
the image by $f_\B$ of an element of $\cyca{\A}$ is  $q$ is proportional to the number of preimages of 
$q$ by $\delta_{u_n}$: it is exactly $\frac1n |\delta_{u_n}^{-1}(\{q\})|$, the probability that a random state is mapped
to $q$ when reading $u_n$. For any word $\omega\in A^*$, let $\PP{\A,\omega}$ be the function from $[n]$ to $[0,1]$ defined by
\begin{equation}\label{eq}
\PP{\A,\omega}(q) = \frac{|\delta_{\omega}^{-1}(\{q\})|}n, \text{ for all }q\in[n].
\end{equation}
From the observations above, we get that once $\A$ is fixed, $f_\B$ is a random $p$-mapping, where
the distribution on $\cyca{\A}$ is given by the restriction of $\PP{\A,u_n}$ to $\cyca{\A}$.

Let $\beta_n=\lfloor n^{\frac14 + 2\epsilon}\rfloor$. Applying Lemma~\ref{lm:p cyclic points} to $f_\B$
yields the following result.

\begin{lem}\label{lm:vn}
Let $\A$ be a fixed automaton of $\E_n$. Consider the random process of building $\B$ by adding a $b$-transition
 to every element of $\cyca{\A}$, choosing the target uniformly and
independently in $[n]$. The  probability that $f_\B$ has more than $\beta_n$ cyclic states or that it is
of height greater than $\beta_n$ is at most $\exp(-n^{\epsilon/4})$. 
\end{lem}

\begin{pf}
Let $c$ denote the number of $\delta_a$-cyclic states in $\A$.
Recall that $f_\B$ is the map from $\cyca{\A}$ to itself
defined by $f_\B(x) = \delta_{bu_n}(x)$, for every $x\in\cyca{\A}$. We consider two cases:

\noindent$\blacktriangleright$ If $c \leq \beta_n$ then there is nothing to prove as $\cyca{\A}$ already
has at most $\beta_n$ states. The probability is therefore $0$ in this case.

\noindent$\blacktriangleright$ If $c > \beta_n$, we can apply Lemma~\ref{lm:p cyclic points}
when $n$ is large enough, since $f_\B$ is a $p$-random mapping on a set
with $c$ elements. Therefore, the  probability
that there are more than $c^{\frac12+\epsilon}$ cyclic points 
or that the height is greater than $c^{\frac12+\epsilon}$ is
at most $\exp(-c^{\epsilon})$. Moreover, observe that since $\A\in\E_{n}$,
\[
c^{\frac12+\epsilon}\leq \alpha_{n}^{\frac12+\epsilon}\leq n^{\frac14 + \epsilon + \epsilon^{2}} \leq \beta_{n}.
\]
This concludes the proof, as $c \geq n^{\frac14}$ gives the announced upper bound.\qed
\end{pf}

For any automaton $\B$ whose $a$-transitions are all defined and whose $b$-transitions starting from
an element of $\cyca{\B}$ are also all defined, let $\cycf{\B}$ denote the set of $f_\B$-cyclic states
of $\B$.

Let $v_n=u_n(bu_n)^{\beta_n}$. 
At this point, 
the number of states to be synchronized 
has been reduced to less than $n^{\frac14+2\epsilon}$ \emph{whp}, since the image of $\delta_{v_n}$ is  included in
$\cycf{\B}$, which has size at most $\beta_n$. It has been achieved by generating all the $a$-transitions, but using 
only the $b$-transitions that start from the $\delta_a$-cyclic states: \emph{whp},
there still are at least 
$n-\alpha_{n}$ unset $b$-transitions that can be used to continue the synchronization.  
Nonetheless, before going on we will first refine the construction of $\B$ introduced in this section
by forbidding some cases, for technical reasons explained in the next section.

\subsection{Forbidding correlated shapes}\label{sec:correlated}
We have  reduced the number of states to be synchronized to no more than
$n^{1/4+2\epsilon}$ states \emph{whp}, but this quantity is still too large for the idea used at the end of the proof, we need to shrink this set once more. 
For an alphabet with at least one more letter $c$, 
we could use the same kind of construction as in Section~\ref{sec:b}, considering the restriction of $\delta_{cv_n}$ to 
the cyclic states of $f_\B$ and would obtain less than roughly $n^{1/8}$ states to be synchronized. This is because
$c$-transitions can be generated independently of what has been done during the previous steps.

Some care is required to adapt this idea for a two-letter alphabet. We aim at using the word $bb$ instead of 
the letter $c$ in the informal description above. Let $\B$ be an incomplete automaton that extends $\A\in\E_n$
and whose defined transitions are all the $a$-transitions and also the $b$-transitions that start from the
$\delta_a$-cyclic states. 
We are interested in building an automaton $\C$ from $\B$, by adding some new random $b$-transitions, in a way such that
$\delta_{bbv_n}$ is totally defined on $\cycf{\B}$. It means that for every $q\in\cycf{\B}$, the state $\delta_b(q)$
must have an outgoing $b$-transition in $\C$. For such an extension $\C$ of $\B$, let $g_\C$ denote the restriction 
of $\delta_{bbv_n}$ to $\cycf{\B}$. 

The main point here is that for a fixed $\B$, we want $g_\C$ to be  defined as a random $p$-mapping, so that we can use Lemma~\ref{lm:p cyclic points} once more. There are, \emph{a priori}, two kind of issues that can prevent this:
\begin{enumerate}
\item When there exists a state $q\in\cycf{\B}$ such that the $b$-transition starting from $\delta_b(q)$ is already defined
in $\B$, that is, when $\delta_b(q)\in\cyca{\B}$.
\item When two  distinct states $q$ and $q'$ in $\cycf{\B}$ are such that $\delta_b(q)=\delta_b(q')$.
\end{enumerate}
Fortunately, the second issue cannot occur: if $\delta_b(q)=\delta_b(q')$ then $f_\B(q)=f_\B(q')$, which is not
possible for two distinct $f_\B$-cyclic states.

The first case can occur, and then the image of $\delta_b(q)$ by $b$ is already defined in $\B$
and therefore $g_\C$ does not follow a $p$-distribution when we build $\C$ by generating the missing
transitions uniformly at random\footnote{Except in the very degenerate case where the restriction
of $\delta_{bb}$ to $\cycf{\B}$ is already a totally defined and constant map in $\B$.}.

On the other hand, if for every $q\in\cycf{\B}$, $\delta_b(q)\notin\cyca{\B}$, then
it is easy to verify that $g_\C$ is a random $p$-mapping: the image of $q\in\cycf{\B}$
by $g_\C$ is a given $x$ when $\delta_{bbv_n}(q)=x$, which is equivalent 
to $\delta_b(\delta_b(q))\in \delta^{-1}_{v_n}(\{x\})$. Since $\delta_b(\delta_b(q))$ is chosen uniformly at
random in $[n]$, it happens with probability $\mathbb{P}_{\B,v_n}(x)$, using the notation of Equation~\eqref{eq}.

We therefore forbid the bad cases and define the set $\F_n$ of incomplete automata $\B$ with $n$ states such that (we add the last condition to what was done in the previous section):
\begin{enumerate}
\item $\B$ extends an element of $\E_n$,
\item the defined transitions of $\B$ are all the $a$-transitions and the $b$-transitions starting from
the states of $\cyca{\B}$,
\item the map $f_\B$  has height at most $\beta_n$ and has at most $\beta_n$ cyclic states,
\item for every $q\in\cycf{\B}$, $\delta_b(q)\notin\cyca{\B}$.
\end{enumerate}

\begin{example}
Assume that for our $\epsilon$, $\beta_{18}=3$.  
The automaton of Example~\ref{ex:b} is in $\F_n$: looking at the mapping $f_\B$, we can see that the $f_\B$-cyclic states are ${\bf 2}$ and ${\bf 13}$, and
their images by $\delta_b$, which are $1$ and $8$ respectively, are not in $\cyca{\B}$. 
The fact that $\delta_b({\bf 3})$ is in $\cyca{\B}$ is not a problem, since ${\bf 3}$ is not a $f_\B$-cyclic state.
\end{example}

If we forget  the last condition in the definition of $\F_n$, 
the other requirements hold \emph{whp}  for every fixed $\A\in\E_{n}$, as a consequence
of Lemma~\ref{lm:vn}. Lemma~\ref{lm:Fn} below states that after our additional 
restriction, we still have a  set large enough.

\begin{lem}\label{lm:Fn}
With high probability a random complete automaton with $n$ states extends an element of $\F_n$. More precisely,
the probability that it does not extend an element of $\F_n$ is at most
$2n^{-1/4+3\epsilon}$, for $n$ large enough.
\end{lem}

\begin{pf}
Fix $\A\in\E_n$, and consider the extensions $\B$ of $\A$ obtained by adding $b$-transitions to the $\delta_a$-cyclic states.
A state $x$ of $\cycf{\B}$ is a \emph{bad state} when there exists $y\in\cycf{\B}$ and $z\in\cyca{\A}$ such that
$y\xrightarrow{b} z$ and $z\xrightarrow{u_n} x$. In such a case, $y$ is the cyclic predecessor of $x$ for the mapping $f_\B$ and
it does not satisfy the last condition of $\F_n$'s definition. Clearly, if $\B$ is not in $\F_n$ then Condition~3 is not satisfied or there is at least one bad state in $\B$

For a given $x\in\cyca{\A}$ and $\ell\in\{0,\ldots,n-1\}$ let us bound from above the probability that $x$ is a bad state and in
a $f_\B$-cycle of length $\ell+1$ when adding the $b$-transitions: 
there must exist $\ell$ distinct states $x_1$, \ldots $x_\ell$ of $\cyca{\A}$, all distinct from $x$, such that
$x\xrightarrow{f_\B} x_1$, $x_1\xrightarrow{f_\B} x_2$, \ldots $x_\ell\xrightarrow{f_\B} x$, and the image by $b$ of $x_\ell$ must be
in $\cyca{\A}$. Hence $\delta_b(x_\ell)$ must belong to $\cyca{\A}\cap\delta_{u_n}^{-1}(\{x\})$. Consequently, the probability that
such a cycle exists is 
\[
\PP{\A,u_n}(x_1)\PP{\A,u_n}(x_2)\cdots\PP{\A,u_n}(x_\ell)\cdot\frac{|\cyca{\A}\cap\delta_{u_n}^{-1}(\{x\})|}n.
\]
We sum this quantity for every possible tuple $(x_1, \ldots, x_\ell)$ of distinct elements of $E_x=[n]\setminus\{x\}$.
We obtain, since there are $\ell!$ ways to order each $\{x_1,\ldots,x_\ell\}$:
\begin{align*}
\ell! \sum_{\substack{x_1<\ldots<x_\ell\\ x_i\in E_x}}
\PP{\A,u_n}(x_1)&\PP{\A,u_n}(x_2)\cdots\PP{\A,u_n}(x_\ell)\cdot\frac{|\cyca{\A}\cap\delta_{u_n}^{-1}(\{x\})|}n
\\ 
& \leq \frac{|\cyca{\A}\cap\delta_{u_n}^{-1}(\{x\})|}n\ell!\binom{n-1}{\ell}\left(\frac{1-\PP{\A,u_n}(x)}{n-1}\right)^\ell,
\end{align*}
by applying Lemma~\ref{lm: bounding sym sums} with $f=\PP{\A,u_n}$, $E=E_x$ and $s = 1 - \PP{\A,u_n}(x)$.
But an easy computation shows that $\ell!\binom{n-1}{\ell}\leq (n-1)^\ell$, hence the probability that $x$ is a bad state in
a $f_\B$-cycle of length $\ell+1$ is at most $\frac1n|\cyca{\A}\cap\delta_{u_n}^{-1}(\{x\})|$.

We now use the union bound and sum the contribution of all $x\in\cyca{\A}$. Since the $\delta_{u_n}^{-1}(\{x\})$ are disjoint, we obtained that the 
probability that there is a bad state in a cycle of length $\ell+1$ is at most
 $\frac1n|\cyc{\A}|$. Hence, since $\A\in\E_n$, this probability is at most than $n^{-1/2+\epsilon}$.

By Lemma~\ref{lm:vn}, the probability that Condition~3 of the definition
of $\F_{n}$ is not satisfied is smaller that $\exp(- n^{\epsilon/4})$. 
Hence, for every fixed $\A\in\E_n$, the probability that there is a bad state or that Condition~3 does not hold is at most
\[
\underbrace{\sum_{\ell = 0}^{\lceil n^{1/4+2\epsilon}\rceil -1} \frac{n^{\frac12+\epsilon}}n}_{\textrm{bad state for typical case}}
+ \underbrace{\exp\left(-n^{\epsilon/4}\right)}_{\textrm{Condition 3 does not hold}}
\leq \frac32\cdot n^{-1/4+3\epsilon}.
\]
Note that we do not need to consider the cases where $\ell+1 > n^{1/4+2\epsilon}$ in the first sum, since they do not satisfy Condition~3.

We therefore obtained a uniform upper bound of $\frac32\cdot n^{-1/4+3\epsilon}$ for every $\A\in\E_{n}$. Since a complete automaton can extend at most one element of $\E_{n}$, the law of total
probabilities applies and yields that: the probability that a complete automaton with $n$ states that extends an element of $\E_{n}$ does not satisfy Condition~3 or Condition~4 is at 
most $\frac32\cdot n^{-1/4+3\epsilon}$, for $n$ large enough. This concludes the proof since  the probability of not being
in $\E_n$ is smaller than $\exp(-n^\epsilon)$. \qed
\end{pf}

\subsection{Adding more random $b$-transitions}
Starting from an element of $\B\in\F_n$, we can now use the idea explained at the beginning of Section~\ref{sec:correlated}, and add
the random $b$-transitions  that are needed for $\delta_{bb}$ to be totally defined on $\cycf{\B}$. 
For such an extension $\C$ of $\B$,
recall that the mapping $g_\C$ is the restriction of $\delta_{bbv_n}$ to $\cycf{\B}$. Let $\cycg{\C}$ denote the
set of $g_\C$-cyclic states in $\C$. Thanks to the last condition of the definition of $\F_n$, we need to randomly
choose the $b$-transitions starting from the images by $\delta_b$ of $\cycf{\B}$, which are all distinct since
two distinct states of  $\cycf{\B}$ cannot have the same image by $\delta_b$.

Let $\gamma_n = \lfloor n^{1/8+4\epsilon}\rfloor$ and let $\G_n$ denote the set of
incomplete automata $\C$ with $n$ states such that:
\begin{enumerate}
\item $\C$ extends an automaton $\B$ of $\F_n$,
\item if $X_\B$ denote the set of images of $\cycf{\B}$ by $\delta_b$, i.e. $X_\B=\{\delta_b(x): x\in\cycf{\B}\}$, 
then the only $b$-transitions of $\C$ are those starting from $\cyca{\B}$ and from $X_\B$;
\item the map $g_\C$ has no more than $\gamma_n$ cyclic states and has height at most $\gamma_n$;
\item for every $q\in\cycg{\C}$ the $b$-transition of $\delta_{bb}(q)$ is undefined.
\end{enumerate}
The last condition in the definition of $\G_n$ is here for the same kind of reasons than the last condition of $\F_n$: 
it is used to ensure some independency for the final step of the synchronization. 


\begin{lem}\label{lm:Gn}
A random complete automaton with $n$ states extends an element of $\G_n$ \emph{whp}. More precisely,
the probability it does not is at most $2n^{-1/4+3\epsilon}$.
\end{lem}

\begin{pf}
We only explain why the last condition  holds \emph{whp}, since one can
easily prove that the probability that Condition~3 does not holds is at most $\exp\left(-n^{\epsilon/8}\right)$
using the same technique as for Lemma~\ref{lm:vn}.

We even prove the stronger result that \emph{whp}, for every 
$q\in\cycf{\B}$ the $b$-transition of $\delta_{bb}(q)$ is  undefined. Once $\C$ is built by adding the needed transitions, 
the defined $b$-transitions start from $\cyca{\B}$ or from $X_\B$. 
Since $\B\in\F_n$, the cardinality of $\cyca{\B}$ and $X_{\B}$ are at most $\alpha_{n}$ and $\beta_{n}$, respectively.
To build $\C$, we iteratively add new random outgoing $b$-transitions for each element of $X_{\B}$; when we add the $i$-th such transition,
the probability it ends in a state that has a defined outgoing $b$-transition is therefore at most $\frac1n(\alpha_{n}+\beta_{n}+i-1)$,
which is smaller than $p_{n}=\frac43 n^{-1/2+\epsilon}$ for $n$ large enough, as $i\leq |X_{\B}|\leq \beta_{n}$. Hence the probability that Condition~4
holds is at least $(1-p_n)^{|X_\B|}$, which is greater than $1-\frac53n^{-1/4+3\epsilon}$ for $n$ large enough, by basic computations.
This conclude the proof after handeling the probability that the other conditions do not hold.\qed
\end{pf}

Let $w_n=v_n(bbv_n)^{\gamma_n}$. Lemma~\ref{lm:Gn} ensures that in a random complete automaton $\A$, the image of  
$\delta_{w_n}$ is included in $\cycg{\A}$, which has size at most $n^{1/8+4\epsilon}$. This concludes
the first part of the synchronization: the word $w_n$  maps the set of states of $\A$
to the much  smaller set of states $\cycg{\A}$ \emph{whp}. 
We will use another technique to finalize the synchronization, which only works because $\cycg{\A}$ is small enough \emph{whp}.

\subsection{Synchronizing the states of $\cycg{\C}$}\label{sec:G}
Let $\lambda_n=\lfloor n^{1/8+5\epsilon}\rfloor$ and let $\C$ be a fixed automaton of $\G_n$.
Starting from $\C\in\G_n$, we now prove that 
the elements of $\cycg{\C}$ can be synchronized \emph{whp}, 
using the remaining randomness of the undefined $b$-transitions. We follow the
idea given at the beginning of Section~\ref{sec:proof} and first
prove that with high enough probability, two states of $\cycg{\C}$ can be synchronized
by a word of the form $b^jw_n$.

\begin{lem}\label{lm:final p q}
Let $\C\in\G_n$ and let $p$ and $q$ be in $\cycg{\C}$. If we add all the missing $b$-transitions
to $\C$ by drawing them uniformly at random and independently, then the probability that
for all $j\in\{0,\ldots, \lambda_n\}$ we have  $\delta_{b^j\cdot w_n}(p)\neq\delta_{b^j\cdot w_n}(q)$
is at most $6n^{-3/8+5\epsilon}$, for $n$ large enough.
\end{lem}

\newcommand{\success}{{\tt success}}
\newcommand{\failure}{{\tt failure}}
\begin{pf}
By definition of $\G_n$, the states $p_2=\delta_{bb}(p)$ and $q_2=\delta_{bb}(q)$ have no outgoing $b$-transitions.
If $p_2=q_2$, then $p$ and $q$ does not satisfy the property for $j=2$. Otherwise we consider
the sequence of pairs of states $(p_i,q_i)$ that is generated 
using the following random process, starting from $i=3$:
\begin{enumerate}
\item generate $(p_{i},q_{i})$ uniformly at random and set $\delta_b(p_{i-1})=p_{i}$ and $\delta_b(q_{i-1})=q_{i}$ in the automaton;
\item if $\delta_{w_n}(p_{i}) = \delta_{w_n}(q_{i})$ 
then stop the process and return a \success\footnote{We call it a \success\  because we have successfully synchronized $p$ and $q$.} (and the property does not hold for $j=i$);
\item otherwise, if $p_i$ or $q_i$ already have an outgoing $b$-transition, stop the process and return a \failure;
\item in other cases, iterate the process for the next value of $i$ by going back to step 1, until $i=\lambda_n$. When $i=\lambda_n$, the
process halts and return a \failure.
\end{enumerate}
Hence we iteratively and in parallel create a sequence of $b$-transitions, starting from $p_2$ and $q_2$. If the process
returns a \success, then clearly the property of the statement does not hold. Thus the probability of returning a \failure\ is an upper bound for the probability that it holds. 

Given that the process has not halted after building up to $(p_{i-1},q_{i-1})$ 
for $1\leq i<\lambda_n$, the probability that it halts at next step and returns a \success\ is
the probability that two randomly chosen elements of $[n]$ have the same image by $\delta_{w_n}$, which is exactly
\[
s_i = \sum_{x\in\cycg{C}}\PP{\C,w_n}(x)^2.
\]
Therefore $s_i\geq \frac1{|\cycg{\C}|}\geq n^{-1/8-4\epsilon}$ by Cauchy-Schwarz inequality.

Let $Y_\C$ denote the set of states of $\C$ that have a defined $b$-transition. Since $\C\in\G_n$,
the only $b$-transitions of $\C$ start from elements
of $\cyca{\C}$ and from their images by $b$, and thus $|Y_\C| \leq 2\alpha_n$.
Hence,  given that the process has not halted after building up to $(p_{i-1},q_{i-1})$ 
for $1\leq i<\lambda_n$, the probability that it halts at next step and returns a \failure\ only depends
on $i$ and satisfies
\[
f_{i} \leq 1-\left(1-\frac{|Y_\C|+2(i-3)}n\right)^2,
\]
because it is not a failure when both $p_i$ and $q_i$ are not in $Y_\C$ and are not one of the $2(i-3)$ states
that get a $b$-transition during the previous iterations of the process. This is an upper bound, since
some cases yield a \success\ (for instance when $p_i=q_i$). Therefore,
\begin{align*}
f_{i} &\leq 1-\left(1-\frac{|Y_\C|+2(i-3)}n\right)^2 = 2\frac{|Y_\C|+2(i-3)}n - \frac{(|Y_\C|+2(i-3))^2}{n^2} \\
& \leq 2\frac{|Y_\C|+2i}n \leq 2\frac{|Y_\C|+2\lambda_n}n \leq 5 n^{-\frac12+\epsilon},
\end{align*}
for $n$ large enough. Note that this bound does not depend on $i$, and
we can therefore bound the probability of a failure given that the process halts when building
$(p_i,q_i)$ by 
\[
\mathbb{P}(\textrm{failure at step }i  \mid \textrm{has not halted before})
= \frac{f_i}{f_i+s_i} \leq \frac{f_i}{s_i} \leq 5 n^{-\frac38+5\epsilon}.
\]

The probability that the process halts at a given step is greater than the probability it halts
and returns a \success. Hence, the probability that the process has build $(p_{\lambda_n},q_{\lambda_n})$ 
without halting
is smaller than or equal to \[ \prod_{i=3}^{\lambda_n} (1-s_i) \leq \left(1-n^{-1/8-4\epsilon}\right)^{\lambda_n-2} = \OO( \exp(-n^{\epsilon})). \]
Putting all together, we get that the probability of a failure is at most
\[
5 n^{-\frac38+5\epsilon} + \OO(\exp(-n^{\epsilon})) \leq 6 n^{-\frac38+5\epsilon},
\]
for $n$ large enough, which concludes the proof.\qed
\end{pf}

To conclude the proof of Theorem~\ref{thm:main}, we use the union bound: for any automaton $\A$ that extends an element
of $\G_n$, which happens \emph{whp}, there are less than $\gamma_n^2$ pairs of states in $\cycg{\A}$; the probability that one of these pairs
$(p,q)$ cannot be synchronized using a word of the form $b^j\cdot w_n$ is therefore at most $\gamma_n^2\cdot4n^{-3/8+5\epsilon}$, which
tends to $0$; more precisely, it is in $\OO(n^{-\frac18+13\epsilon})$. 

To obtain the length of the synchronizing word, we apply Lemma~\ref{lm:pairs} to the elements
of $\cycg{\A}$: \emph{whp} there are at most $\gamma_{n}$ such states, which can be pairwise
synchronized using  words of the form $b^{j}w_{n}$, of length at most $|w_{n}|+\lambda_{n}$. 
Hence, the set $\cycg{\A}$ can be synchronized using a word $z$ of length at most
$(\gamma_{n}-1)(|w_{n}|+\lambda_{n})$, which is asymptotically equivalent to $n^{1+11\epsilon}$.
To conclude, observe that $w_{n}z$ is synchronizing, and also of length asymptotically equivalent to $n^{1+11\epsilon}$.
Changing $\epsilon$ into $\epsilon/13$ yields the result.

\section{Conclusion}

In this article we  proved that most complete automata are synchronizing, since they admit a synchronizing word of length smaller than
$n^{{1+\epsilon}}$ with high probability. 

Note that our proof can be turned into a probabilistic algorithm to try to quickly find
a synchronizing word:
Compute the action of 
$\delta_{u_{n}}$,  $\delta_{v_{n}}$ and then $\delta_{w_{n}}$ in linear time. Once it is done, 
check whether the property of Lemma~\ref{lm:final p q} holds for every pair of elements of the image of $\delta_{w_{n}}$, which is small with high probability.
Experiments seem to indicate that the algorithm behaves way better in practice than its theoretical
analysis: it looks like an important proportion of automata that fail to fulfill every step of our construction are
still detected as synchronizing by the combination of computing $\delta_{w_n}$ and synchronizing the states of its image with the $b^j$'s.

A natural continuation of this work 
is to prove that with high probability automata are synchronized by  words that are way shorter
than $n^{1+\epsilon}$. Experiments have been done~\cite{Kisielewicz}, 
and seem to indicate that the expected length of the
smallest synchronizing word is often sublinear, probably in $\sqrt{n}$. There is plenty of room to improve our construction, as the synchronizing
words we obtain have very specific shapes, but it might be quite difficult to have a proof that matches what was observed during the experiments of~\cite{Kisielewicz}.


\para{Acknowledgments:} the author would like to thank Marie-Pierre B\'eal and Dominique
Perrin for their interest in this work since the very beginning.

\bibliographystyle{abbrv}
\bibliography{biblio}

\begin{thebibliography}{10}

\bibitem{aldous}
D.~Aldous, G.~Miermont, and J.~Pitman.
\newblock Brownian bridge asymptotics for random p-mappings.
\newblock {\em Electron. J. Probab}, 9:37--56, 2004.

\bibitem{berlinkov}
M.~V. Berlinkov.
\newblock On the probability to be synchronizable.
\newblock {\em arXiv}, abs/1304.5774, 2013.
\newblock http://arxiv.org/abs/1304.5774.

\bibitem{cameron}
P.~J. Cameron.
\newblock Dixon's theorem and random synchronization.
\newblock {\em Discrete Mathematics}, 313(11):1233--1236, 2013.

\bibitem{rm}
P.~Flajolet and A.~M. Odlyzko.
\newblock Random mapping statistics.
\newblock In {\em EUROCRYPT}, volume 434 of {\em LNCS}, pages 329--354.
  Springer, 1989.

\bibitem{FSBook}
P.~Flajolet and R.~Sedgewick.
\newblock {\em Analytic Combinatorics}.
\newblock Cambridge University Press, 2009.

\bibitem{frankl}
P.~Frankl.
\newblock An extremal problem for two families of sets.
\newblock {\em Eur. J. Comb.}, 3:125--127, 1982.

\bibitem{harris}
B.~Harris.
\newblock Probability distributions related to random mappings.
\newblock {\em The Annals of Mathematical Statistics}, 31(4):1045--1062, 1960.

\bibitem{Karp90}
R.~M. Karp.
\newblock The transitive closure of a random digraph.
\newblock {\em Random Struct. Algorithms}, 1(1):73--94, 1990.

\bibitem{Kisielewicz}
A.~Kisielewicz, J.~Kowalski, and M.~Szykula.
\newblock A fast algorithm finding the shortest reset words.
\newblock In D.-Z. Du and G.~Zhang, editors, {\em COCOON}, volume 7936 of {\em
  Lecture Notes in Computer Science}, pages 182--196. Springer, 2013.

\bibitem{kolcin}
V.~Kol{\v{c}}in.
\newblock {\em Random Mappings: Translation Series in Mathematics and
  Engineering}.
\newblock Translations series in mathematics and engineering. Springer London,
  Limited, 1986.

\bibitem{Olschewski}
J.~Olschewski and M.~Ummels.
\newblock The complexity of finding reset words in finite automata.
\newblock In P.~Hlinen{\'y} and A.~Kucera, editors, {\em MFCS}, volume 6281 of
  {\em Lecture Notes in Computer Science}, pages 568--579. Springer, 2010.

\bibitem{pin}
J.-E. Pin.
\newblock On two combinatorial problems arising from automata thery.
\newblock {\em Annals of Discrete Mathematics}, 17:535--548, 1983.

\bibitem{cerny}
J.~\v{C}ern\'y.
\newblock Pozn\'amka k. homog\'ennym experimentom s konecnymi automatmi.
\newblock {\em Matematicko-fyzikalny \v{C}asopis Slovensk}, 14, 1964.

\bibitem{volkov08}
M.~V. Volkov.
\newblock Synchronizing automata and the \v{C}ern\'y conjecture.
\newblock In {\em LATA'08}, volume 5196 of {\em LNCS}, pages 11--27. Springer,
  2008.

\end{thebibliography}

\end{document}